\documentclass[12pt,preprint]{aastex}
\shorttitle{Mergers and sdB stars}
\newcommand{\Msun}{\rm\,M_\odot}

\usepackage[usenames,dvips]{color}
\usepackage{graphicx}

\begin{document}

\title{Common Envelope Mergers: A Possible Channel for Forming Single sdB Stars}

\author{Michael Politano\altaffilmark{1}, Ronald E. Taam\altaffilmark{2,3,4}, Marc van der Sluys\altaffilmark{2}, and Bart Willems\altaffilmark{2}}

\altaffiltext{1}{Department of Physics, Marquette University, P.O. Box 1881, Milwaukee, WI 53201-1881}

\altaffiltext{2}{Department of Physics and Astronomy, Northwestern University, 2131 Tech Drive, Evanston, IL 60208}

\altaffiltext{3}{Academia Sinica Institute of Astronomy \& Astrophysics - National Tsing Hua University - TIARA, Hsinchu, Taiwan}

\altaffiltext{4}{Academia Sinica Institute of Astronomy \& Astrophysics - TIARA, Taipei, Taiwan}

\begin{abstract}

We quantify an evolutionary channel for single sdB stars based on mergers of binaries containing a red giant star and a lower mass main sequence or brown dwarf companion in our Galaxy.  Population synthesis calculations that follow mergers during the common envelope phase of evolution of such systems reveal a population of rapidly rotating horizontal branch stars with a distribution of core masses between $0.32\Msun - 0.7\Msun$ that is strongly peaked between $0.47\Msun - 0.54 \Msun$. The high rotation rates in these stars are a natural consequence of the orbital angular momentum deposition during the merger and the subsequent stellar contraction of the merged object from the tip of the red giant branch.  We suggest that centrifugally enhanced mass loss facilitated by the rapid rotation of these stars may lead to the formation of single sdB stars for some of these objects.

\end{abstract}


\section{Introduction}

Subdwarf B (sdB) stars are core He-burning stars that have an extremely thin ($\lesssim$ 0.02$\Msun$) hydrogen-rich envelope (for recent reviews, see \citealp{heb08,cat07,mon07}).  Observations indicate that 40 -- 70\% of sdB stars in the field are in binaries \citep{max01,ree04,nap04,lis05,mor06}.

\citet{han02,han03} identified and performed extensive population synthesis calculations of two channels for the formation of sdB stars in binaries in the field: (1) one or two phases of unstable Roche lobe (RL) overflow leading to common envelope (CE) evolution and (2) one or two phases of stable RL overflow, and one channel for the formation of single sdB stars in the field:  the merger of two He white dwarfs (WDs).  Their model of the population of sdB stars in binaries has been very successful in explaining several features of the observed field population, such as the prevalence of systems with $P \lesssim$ 10 days, and this model is generally well accepted.   Their model of the single sdB star population in the field predicts a relatively flat distribution of masses ranging from $\sim 0.4\Msun - 0.7\Msun$ (see Fig.\,12 in \citealp{han03}).   This distribution differs from the mass distribution expected from single star horizontal branch (HB) models (e.g., \citealp{dor93}), which predict a narrow distribution of sdB star masses strongly peaked at $\sim0.47\,\Msun$.  Observationally, masses for seven single field sdB stars have been determined using asteroseismology (see \citealp{ran07} and references therein).  Six are very close to 0.47$\Msun$ and one is 0.39$\Msun$.   However, it is not evident whether these masses are representative of the single sdB star population as a whole, since only $\sim$ 5\% of sdB stars pulsate.  Further, reliable mass determinations using asteroseismology depend on correctly identifying the pulsational modes of the star, which can be quite difficult.  

The formation of single sdB stars also has been modeled using channels that do not require the merger of two WDs, but rather involve a single star with a
phase of enhanced mass loss on the red giant branch (RGB).  Various suggestions for the cause of this enhanced mass loss include He mixing driven by internal rotation (Sweigart 1997), spin up of the primary via interactions with a planet in a close orbit \citep{sok98}, and the existence of a sub-population of ZAMS stars with a high He-abundance (e.g., D'Antona et al. 2005).   In all of these proposed causes, the amount of mass loss on the RGB must be fine-tuned in order to yield the very small envelope masses observed in sdB stars.  However, it is not evident how such fine-tuning is incorporated naturally into the physical mechanisms that have been suggested to enhance the mass loss.

In this Letter, we quantify a channel for the formation of single sdB stars in the field of our Galaxy: the 
evolution of objects resulting from the merger of stellar or sub-stellar objects with RGB stars during a phase of CE evolution initiated by unstable RL overflow or by a tidal instability (hereafter, referred to as the "CE merger channel").  We carry out detailed population synthesis calculations of these mergers and predict the resulting population of HB stars at the present epoch, some of which may have very small envelope masses and could be observed as single sdB stars.  Although our treatment was developed independently, the idea that single sdB stars might result from the merger of an RGB star and a low mass companion during CE evolution was suggested by \citet{sok98} and explored further by \citet{sok00,sok07}.

\section{Method}

We use the Monte Carlo population synthesis code developed by Politano \citep{pol96,pw07}.   For the present study, two major modifications have been incorporated: 1)  the analytic fits that had been used previously (see \citealp{pol96}) have been replaced by numerical tables of 116 up-to-date stellar models ranging in mass from 0.5 to 10.0$\Msun$ in increments of 0.1$\Msun$ and from 10.5 to 20$\Msun$ in increments of 0.5$\Msun$ and 2) tidal effects which act to synchronize the rotation of the primary with the orbit.  

The updated stellar models were calculated using a version of the binary stellar evolution code, {\it STARS}, developed by Eggleton \citep{1971MNRAS.151..351E,1972MNRAS.156..361E,2002ApJ...575..461E} and updated as described in \citet{1995MNRAS.274..964P}.   The treatment of convective mixing, convective overshooting, the helium flash and our definition of the core mass and of the envelope binding energy\footnote{The gravitational potential and internal thermal energies of the envelope are included in the binding energy, but not the ionization/recombination energy.} are given in \cite{2006A&A...460..209V}.    Convective overshooting on the main sequence is taken into account for stars with $M \geq1.2\Msun$ and helium is ignited degenerately in stars with $M\leq2.0\Msun$.  The initial composition of our model stars is similar to solar composition: $X=0.70, Y=0.28$ and $Z = 0.02$.  Mass loss via stellar winds is incorporated in the models using a Reimers-type prescription \citep{rei75} with an effective $\eta$ of 0.2 and the prescription by \cite{1988A&AS...72..259D}, which is negligible except for the most massive stars in our grid. 

The second modification to the code is that synchronism of the primary's rotation with the orbit is now assumed when the primary is on the RGB or the asymptotic giant branch (AGB).  For these cases, after each time step, the new total angular momentum (AM) of the system is calculated, accounting for losses due to stellar winds.  Synchronism is then enforced by redistributing this total AM such that the rotational angular velocity of the primary is equal to the orbital angular velocity of the binary.   In all cases, we assume a circular orbit and rigid body rotation for the primary and we neglect the rotational AM of the secondary.  We also assume that the rotational AM of the primary on the ZAMS is negligibly small. 
 
In our population synthesis calculations, we begin with 10$^7$ zero-age main sequence (ZAMS) binaries.  We assume that these binaries are distributed over primary mass, $M_\mathrm{p,0}$, according to a \citet{mil79} IMF, over orbital period uniformly in log $P_0$ \citep{abt83} and over mass ratios uniformly in $q_0$ (i.e., $g(q_0)\,dq_0 = 1\,dq_0$), where $q_0$ = $M_\mathrm{s,0}$/$M_\mathrm{p,0}$ and $M_\mathrm{s,0}$ is the mass of the secondary star \citep{duq91,maz92,gol03}.  Once the primary mass and mass ratio have been generated for a given binary, the secondary mass is determined by their product:  $M_\mathrm{s,0}$ = $q_0\,M_\mathrm{p,0}$.  We adopt a minimum primary mass of 0.95$\Msun$, a maximum primary mass of 10$\Msun$ and a minimum secondary mass of 0.013$\Msun$ in our calculations.  For secondaries with masses less than 0.5$\Msun$, including substellar secondaries, we use detailed stellar models from the Lyon group (e.g., \citealp{bar03} and refs. therein).  For secondaries with masses greater than or equal to 0.5$\Msun$, we adopt the same stellar models used for the primaries.   We assume that the duration of the CE phase is negligible compared to the other time scales (e.g., \citealp{ibe93,taa00}), that the age of the Galaxy is 10$^{10}$ yrs, and that the star formation rate throughout the Galaxy's history has remained constant. 

To model the population of mergers between giant primaries and less massive companions, we employ the following scenario.   A given ZAMS binary is evolved until the primary reaches the base of the RGB, at which point we assume synchronism between the primary's rotation and the orbit.   As the primary ascends the RGB, CE evolution can be initiated either via unstable RL overflow or a tidal instability.  We adopt the criterion for unstable mass transfer via RL overflow given by eq. 57 in \citet{hur02}.  To determine if merger occurs within the CE, simple energy considerations are used to relate the pre- and post-CE orbital separations \citep{tut79} according to a standard prescription (e.g., \citealp{wil05}).  The binding energy of the primary's envelope at the onset of the CE phase is determined directly from the stellar models.   The efficiency at which orbital energy is transferred to the CE, parameterized by $\alpha_\mathrm{CE}$, embodies a major uncertainty in this simplified prescription.  We have chosen $\alpha_\mathrm{CE}$ = 1 as our standard model and we investigate the dependence of our results on $\alpha_\mathrm{CE}$.  Merger is assumed to have occurred if the radius of the secondary is larger than its RL radius at the end of the CE phase. In addition, we consider mergers that result from a tidal instability.  If the moment of inertia of the primary exceeds one-third of the moment of inertia of the binary before the primary fills its RL on the RGB or AGB, there will no longer be sufficient AM within the orbit to keep the envelope rotating synchronously \citep{dar79} and the orbit will continue to shrink until the envelope is ejected or a merger occurs. 

Available detailed hydrodynamical models of stellar mergers have focused on high-velocity collisions between stars and are more appropriate for mergers that occur as a result of stellar interactions within dense stellar environments, such as at the center of clusters (e.g., \citealp{sil97,sil01,lom02}).  In the absence of detailed models that are applicable to stellar mergers during CE evolution, we use a very simplified model for the merger process.  The key assumptions in our model are listed below. 

1. In most cases, the merger of the entire secondary with the primary leads to an object spinning at several times its break-up rotational velocity, given by $v_\mathrm{br} =  (GM/R)^{1/2}$, where $M$ and $R$ are the object's mass and radius, respectively.  Consequently, we assume that a rapid mass loss phase occurs when the rotational velocity of the object exceeds a critical rotational velocity, $v_\mathrm{crit} = \onethird v_\mathrm{br}$, which corresponds to centrifugal forces contributing about 10\% of the force against gravity.  

2. We assume that the primary assimilates only as much mass from the secondary as is required such that it rotates at this critical velocity and we assume that the remaining secondary mass is expelled from the system. 

3. Although the merged object is not likely to be in a state of thermal equilibrium initially, we assume for definiteness that the radius of the merged object corresponds to that of a star in the model grid for the same core mass and total mass after it achieves thermal equilibrium.

4. In cases where choosing a model with the same core mass and total mass would result in an unphysical evolutionary regression (such as an RGB star reverting to a Hertzsprung gap star as a result of the merger), we assume that the radius of the merged object corresponds to a star in the model grid with the same total mass, but with the smallest core mass within the same evolutionary state as prior to merger.

The subsequent evolution of the merged object, including its rotational velocity, $v_\mathrm{rot}$, is then followed until the present epoch. If $v_\mathrm{rot}>v_\mathrm{crit}$ at any time, we assume that AM is removed via mass loss until the object is rotating at exactly $v_\mathrm{crit}$.  We make the same assumptions regarding the choice of stellar model following mass loss as we did following mass accretion during merger.  If the merged object completes its evolution as an AGB star before the present epoch is reached, we assume the merged object will be a white dwarf at the present epoch.

\section{Results}

In Figure\,1, we show the two-dimensional distribution of the rotational velocities (as a fraction of $v_\mathrm{crit}$) and the envelope masses ($M_\mathrm{env}$) for our predicted population of stars on the HB at the present epoch.  The distribution of rotational velocities rises slowly from 0.08$\,v_\mathrm{crit}$ to $\sim 0.35\,v_\mathrm{crit}$, remains fairly uniform up to 0.95$\,v_\mathrm{crit}$, and then rises rapidly, with 28\% of the population rotating at $v_\mathrm{crit}$.   The distribution of envelope masses peaks at $\sim$$\,$0.35$\Msun$ and decreases with increasing envelope mass, following the IMF for the primaries.  Approximately 24\% of the population has $M_\mathrm{env}<0.50\,\Msun$ and there are a small number of objects with very low envelope masses, ranging from 0.27$\Msun$ down to 0.07$\Msun$.

In Figure\,2, we show the distribution of core masses for our predicted population of HB stars at the present epoch (solid curve).  We find that the core masses range from $0.32\,\Msun$ to $\sim 0.7\,\Msun$ and the distribution is strongly peaked between 0.47 and 0.54$\Msun$, with 78\% of the population having core masses in this range.  For comparison, the dashed curve in Figure\,2 shows the predicted distribution of core masses in a present-day population of HB stars that results from normal single star evolution.  The two distributions are rather similar, although the dashed curve is slightly less peaked than the solid curve, with 67\% of the dashed population having core masses between 0.47 and 0.54$\Msun$.  Also, the number of present-day HB stars from our merger channel is about an order of magnitude smaller than the number predicted from normal single star evolution and the HB stars from our merger channel are rotating 10 -- 30 times faster on the average than those from normal single star evolution.

All of the HB stars in Fig.\,1 are the result of a merger that occurred when the primary was on the RGB.  During the merger, we have assumed that the primary ceases to assimilate mass from the secondary once the rotational velocity of the merged object reaches a critical rotational velocity, $v_\mathrm{crit} = \onethird\,v_\mathrm{br}$.  Thus, by assumption, the merged object begins its evolution to the present epoch rotating at $v_\mathrm{crit}$.  As the object evolves up the RGB, the increase in radius causes the moment of inertia of the envelope to increase and the rotational velocity to decrease correspondingly. The rotational velocity continues to decrease until the object reaches the tip of the RGB and He is ignited in the core.  If He is ignited degenerately, the object's radius decreases by a factor of  5 -- 15 and the moment of inertia of the envelope decreases dramatically.  If He is ignited non-degenerately, the object's radius and moment of inertia still decrease, but much less dramatically.   In the majority of cases, the decrease in the moment of inertia causes the object's rotational velocity to exceed $v_\mathrm{crit}$.  Should this occur, we artificially remove mass from the envelope (at the specific AM of envelope's radius) before taking the next evolutionary time step, until the envelope is just rotating at $v_\mathrm{crit}$.  In approximately three-quarters of the population, the growth of the object's radius during core He burning is sufficient to eventually reduce the rotational velocity of the envelope to sub-critical by the present epoch.  However, in the remaining one-quarter of the population, the envelope continues to rotate at $v_\mathrm{crit}$ at the present epoch and rotationally enhanced mass loss may continue.  

We have investigated the sensitivity of these results to our assumed choices for $\alpha_\mathrm{CE}$ and $g(q_0)$ by selecting two other choices for each:  $\alpha_\mathrm{CE}=0.5$ and 0.1, and $g(q_0)=q_0$ and $q_0^{-0.9}$.  Selected results for these cases and for our standard model are given in Table 1.  We find that the various features of the population described above do not depend significantly on the assumed choice for either $\alpha_\mathrm{CE}$ or $g(q_0)$.  The distribution of core masses remains strongly peaked, with the fraction between 0.47$\Msun$ and 0.54$\Msun$ varying only slightly from 78\% to 84\%.  A significant fraction of the population still rotates at $v_\mathrm{crit}$, with this fraction varying from 29\% to 39\%, and has envelope masses less than 0.5$\Msun$, with this fraction varying from 24\% to 29\%.  In all models, a small number of systems have very low envelope masses, down to 0.07 -- 0.08 $\Msun$.   The choices for $\alpha_\mathrm{CE}$ and $g(q_0)$ have a stronger effect on the total number of stars in the population, $N$, as indicated by the last column in Table 1.  Compared with our standard model, $N$ increases by as much as 50\% for the $\alpha_\mathrm{CE}=0.1$ model and decreases by 56\% for the $g(q_0)=q_0^{-0.9}$ model.

We have also investigated the effect of increasing the assumed value of $v_\mathrm{crit}$ on our population of HB stars.  As shown in Table 1, choosing $v_\mathrm{crit}=v_\mathrm{br}$ still results in a significant fraction of the population rotating critically and in a sharply peaked distribution of core masses.  However, the fraction of the population with $M_\mathrm{env}<0.5\Msun$ is drastically reduced, resulting in no systems with $M_\mathrm{env}<0.35\Msun$.  As this choice for $v_\mathrm{crit}$ is rather extreme, our results suggest that a critical velocity corresponding to a fraction of the break-up velocity is more appropriate.

\section{Discussion}

In our model population of HB stars, a small fraction ($\la 0.1$\%) have envelopes with masses less than 0.1$\Msun$ (see Fig.\,1).  Approximately one-half of these systems have rotational velocities equal to $v_\mathrm{crit}$, suggesting that these stars will continue to experience enhanced mass loss.  Follow-up evolutionary calculations of core-He burning stars with an initial envelope mass of 0.1$\Msun$ (using the same {\it STARS} stellar evolution code with an artificially large wind to simulate rapid mass loss) suggest that at some point ($\sim0.07\Msun$) the star will begin to contract upon further mass loss and move blueward along the HB.  It is conceivable, therefore, that these critically-rotating, very low envelope mass HB stars in our population may lose sufficient envelope material to become sdB stars.

These potential sdB systems constitute a very small fraction of our population.  Upon examination of Fig.\,1, this fraction would increase substantially if the envelope masses in our population were systematically reduced by $\sim0.2$$\Msun$.  Such a reduction might be accomplished by assuming a higher rate of mass loss due to stellar winds on the RGB.   We have adopted a standard Reimers-like prescription for mass loss in our stellar models, but it has been suggested that a rapidly rotating envelope may lead to enhanced mass loss on the RGB (e.g., \citealp{sw97}).  In addition, we find that our assumption of synchronization on the RGB results in small number of primaries being spun up to $v_\mathrm{crit}$ prior to contacting their RL.  This number is increased significantly if we choose $v_\mathrm{crit}<\onethird\,v_\mathrm{br}$.  For example, if $v_\mathrm{crit}=0.1\,v_\mathrm{br}$, two-thirds of the primaries are spun up to $v_\mathrm{crit}$ prior to contacting their RL.  However, we are not able to properly model either possibility in our population synthesis calculations, since a self-consistent treatment of mass loss in such cases would require stellar evolutionary models in which mass loss via winds is coupled to the rotation and this is beyond the current capabilities of our code.  

Reliable quantitative predictions concerning the absolute number of single sdB stars produced from our CE merger channel are not provided here, since our treatment of the merger process is too crude and is severely limited by the lack of models of stellar mergers relevant to CE evolution.  Rather, our intent in this Letter has been to investigate a possible evolutionary channel for the formation of single sdB stars and to motivate its further study.  In comparison with previously suggested single sdB star channels, this channel has certain advantages:   (1) the AM necessary to spin up the envelope prior to the HB is explicitly and naturally provided by the orbital AM of the companion (see also \citealp{sok98}); (2) the CE phase may be initiated at any point along the RGB, not just near the tip, since mass loss from the envelope also occurs on the HB as a result of super-critical rotation caused by the star's contraction following He ignition in the core; and (3) the predicted spectrum of core masses is consistent with that expected in HB stars.

Further, our population synthesis calculations are, to the best of our knowledge, the first to explore in any substantive manner a possible observational counterpart to mergers resulting from CE evolution.  It is generally accepted that mergers are likely to occur during CE evolution (e.g., \citealp{ibe93,taa00}) and with a frequency roughly comparable to that of the surviving post-CE binaries (e.g., \citealp{pw07}).  Thus, there may exist a significant, but as yet unrecognized, population of objects in our Galaxy that are the result of a merger during CE evolution.  Assuming this is true, identification of possible observational counterparts of these systems is likely to stimulate the development of models necessary to study them.

\acknowledgements{
  We thank P.\ Eggleton for making his binary evolution code available to us, and the referee, Z.\ Han, for his useful suggestions to improve the paper.  M.P.\ acknowledges funding from the Wisconsin Space Grant Consortium to Marquette University.  R.T.\ acknowledges support from NSF AST-0703950 to Northwestern University and from the Institute of Astronomy and Astrophysics of the Academia Sinica in Taipei, Taiwan.  M.V.\ and B.W.\ acknowledge support from NSF CAREER Award AST-0449558 and NASA BEFS NNG06GH87G, respectively, to Northwestern University.
}

\clearpage

\begin{deluxetable}{lcccc}
\tablecolumns{5}
\tablecaption{Selected results for various input parameters}
\tablehead{
\colhead{Model} & \multicolumn{3}{c}{Fraction of population with} & \colhead{$N/N_\mathrm{stand}$} \\
\cline{2-4}
\colhead{} & \colhead{$0.47\Msun \leq$ $M_\mathrm{c}$} & \colhead{$v_\mathrm{rot} =$} & \colhead{$M_\mathrm{env}<$} \\
\colhead{} & \colhead{$\leq 0.54\Msun$} & \colhead{$v_\mathrm{crit}$} & \colhead{$0.5\Msun$}  
}
\tablewidth{0pt}
\startdata
standard & 0.78 & 0.28 & 0.24 & 1.00 \\
$\alpha_\mathrm{CE} = 0.5$ & 0.81 & 0.32 & 0.25 & 1.19 \\
$\alpha_\mathrm{CE} = 0.1$ & 0.84 & 0.39 & 0.29 & 1.50 \\
$g(q_0)=q_0$ & 0.79 & 0.29 & 0.24 & 1.01 \\
$g(q_0)=q_0^{-0.9}$ & 0.78 & 0.31 & 0.24 & 0.44 \\
$v_\mathrm{crit}=v_\mathrm{br}$ & 0.66 & 0.21 & 0.06 & 1.10  
\enddata
\end{deluxetable}

\clearpage

\begin{figure}
\begin{center}
\resizebox{8.0cm}{5.0cm}{\includegraphics{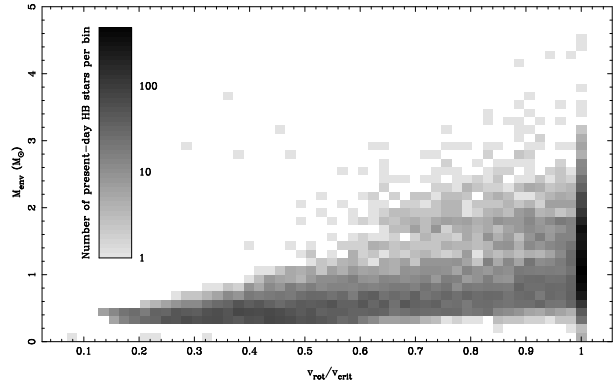}}
\caption{Unnormalized two-dimensional distribution of envelope masses and rotational velocities in our model population of present-day HB stars from the CE merger channel.  The rotational velocities are given as a fraction of the assumed critical rotational velocity for mass loss, $v_\mathrm{crit}$ (see text).  The total number of HB stars in our present-day model population, $N$, is 13,100.}  
\end{center}
\end{figure}

\clearpage

\begin{figure}
\begin{center}
\resizebox{8.0cm}{5.0cm}{\includegraphics{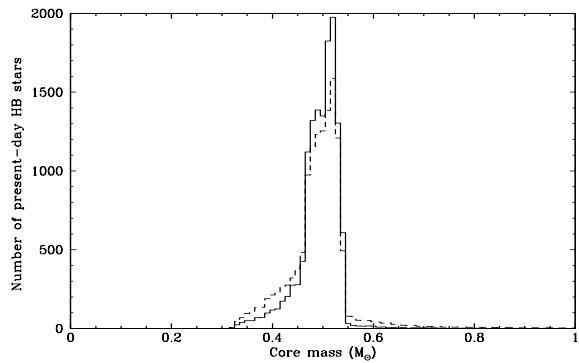}}
\caption{Unnormalized distribution of core masses in our model population of present-day HB stars that results from the CE merger channel (solid curve).  The distribution of core masses in a model population of present-day HB stars that results from standard single star evolution is also shown for comparison (dashed curve).  The dashed population was calculated using the same stellar evolution models, IMF, star formation rate and total number of ZAMS stars as in our mergers population.  The total number of HB stars shown in the dashed curve has been normalized to the same number as in the solid curve to facilitate comparison; in actuality $N_\mathrm{sing}=8.5\,N_\mathrm{merg}$.} 
\end{center}
\end{figure}

\end{document}